\documentclass[lettersize,journal]{IEEEtran}
\usepackage{amsmath,amsfonts,amssymb}
\usepackage{algorithmic}
\usepackage{algorithm}
\usepackage{array}
\usepackage[caption=false,font=footnotesize,labelfont=rm,textfont=rm]{subfig}
\usepackage{textcomp}
\usepackage{stfloats}
\usepackage{url}
\usepackage{verbatim}
\usepackage{graphicx}
\usepackage{cite}
\usepackage{color}
\begin{document}

\title{Near-Field Multiuser Communications Aided by Movable Antennas}

\author{Jingze Ding, \IEEEmembership{Graduate Student Member, IEEE},
		Lipeng Zhu, \IEEEmembership{Member, IEEE},
		Zijian Zhou, \IEEEmembership{Member, IEEE}, \\
		Bingli Jiao, \IEEEmembership{Senior Member, IEEE},
		and Rui Zhang, \IEEEmembership{Fellow, IEEE}
        % <-this % stops a space
\thanks{This work was supported by National Natural Science Foundation of China under Grant 62171006.  The calculations were supported by High-Performance Computing Platform of Peking University. \textit{(Corresponding authors: Jingze Ding; Zijian Zhou.)}}
\thanks{J. Ding and B. Jiao are with School of Electronics, Peking University, Beijing 100871, China (e-mail: djz@stu.pku.edu.cn; jiaobl@pku.edu.cn).}
\thanks{L. Zhu is with the Department of Electrical and Computer Engineering, National University of Singapore, Singapore 117583 (e-mail: zhulp@nus.edu.sg).}
\thanks{Z. Zhou is with School of Science and Engineering, The Chinese University of Hong Kong, Shenzhen, Guangdong 518172, China (e-mail: zjzhou@cuhk.edu.cn).}
\thanks{R. Zhang is with School of Science and Engineering, Shenzhen Research Institute of Big Data, The Chinese University of Hong Kong, Shenzhen, Guangdong 518172, China (e-mail: rzhang@cuhk.edu.cn). He is also with the Department of Electrical and Computer Engineering, National University of Singapore, Singapore 117583 (e-mail: elezhang@nus.edu.sg).}
}



\maketitle

\begin{abstract}
This letter investigates movable antenna (MA)-aided downlink (DL) multiuser communication systems under the near-field channel condition, where both the base station (BS) and the users are equipped with MAs to fully exploit the degrees of freedom (DoFs) in antenna position optimization. We develop a general channel model to accurately describe the channel characteristics in the near-field region and formulate an MA-position optimization problem to minimize the BS's transmit power subject to users' individual rate constraints. To solve this problem, we propose a two-loop dynamic neighborhood pruning particle swarm optimization (DNPPSO) algorithm that significantly reduces the computational complexity as compared to the standard particle swarm optimization (PSO) algorithm while achieving similar performance. Simulation results validate the effectiveness and advantages of the proposed scheme in power-saving for near-field multiuser communications.
\end{abstract}


\section{Introduction}
\IEEEPARstart{O}{ver} the past few decades, multiuser multiple-input multiple-output (MU-MIMO) systems have garnered significant attention and been widely implemented in existing wireless communication systems \cite{MU-MIMO}. However, conventional MU-MIMO systems predominantly adopt fixed-position antennas (FPAs), inherently limiting the exploitation of spatial degrees of freedom (DoFs) because the channel variation in the continuous spatial field is not fully exploited. Therefore, movable antenna (MA) technology has been proposed to address these limitations, introducing a new paradigm in MU-MIMO system design \cite{MA1}. Specifically, MAs are connected to the radio frequency chains via flexible cables and can be dynamically repositioned with the assistance of mechanical drivers \cite{MA2}, which is also known as fluid antenna system with other implementation ways in antenna positioning \cite{FAS}. This innovation has the potential to actively reconfigure channels and thereby significantly enhance communication performance by fully utilizing the spatial DoFs. Recent studies have demonstrated the feasibility and benefits of MA-aided systems \cite{Ding1,Ding2,channel_est1}.

In practical applications, due to the need to accommodate the free movements of multiple MAs and further improve communication performance, MA-aided MU-MIMO systems typically have larger aperture sizes compared to conventional FPA-based systems \cite{MA_near}. Additionally, to attain broader bandwidth, wireless communication systems are increasingly operating at higher frequency bands \cite{near2}. As a result, the increase in Rayleigh distance invalidates the far-field plane-wave model and necessitates the near-field spherical-wave model. However, the previously proposed field-response channel model for MA is based on the far-field assumption \cite{MA1}. Research on MA-aided systems under the near-field channel condition is still in its infancy \cite{MA_near,near3,near4}.

Motivated by the aforementioned discussions, we develop a near-field multiuser downlink (DL) communication system in this letter, where both the base station (BS) and the users are equipped with MAs to fully exploit the spatial DoFs in antenna position optimization in large-size regions. We consider the practical non-parallelism between the transmit and receive regions and develop a general near-field channel model for the MA-aided multiuser system under the point scatterer assumption. Subsequently, we propose a two-loop dynamic neighborhood pruning particle swarm optimization (DNPPSO) algorithm to minimize the transmit power by jointly optimizing the beamformers and the MA positions. Finally, simulation results demonstrate the effectiveness of the proposed algorithm and validate the advantages of MA in exploiting near-field channel spatial variation gain for saving the transmit power.

\textit{Notation:} $a/A$, $\mathbf{a}$, $\mathbf{A}$, and $\mathcal{A}$ denote a scalar, a vector, a matrix, and a set, respectively. ${\left(  \cdot  \right)^{T}}$, ${\left(  \cdot  \right)^{H}}$, and $\left\|  \cdot  \right\|_2$ denote the transpose, conjugate transpose, and Euclidean norm, respectively. $\odot$ represents the Hadamard product. $\mathbf{0}_N$ denotes an $N$-dimensional vector with all elements equal to 0. $\mathbf{I}_N$ denotes an identical matrix of size $N \times N$. $\mathbb{C}^{M \times N}$ and $\mathbb{R}^{M \times N}$ denote the sets for complex and real matrices of $M \times N$ dimensions, respectively. $\mathcal{A} \backslash \mathcal{B}$ represents the subtraction of set $\mathcal{B}$ from set $\mathcal{A}$. $\triangleq$ stands for ``defined as''.
\section{System Model}\label{2}
\begin{figure}
	\centering
	\subfloat[]{\label{system_model}\includegraphics[width=1\columnwidth]{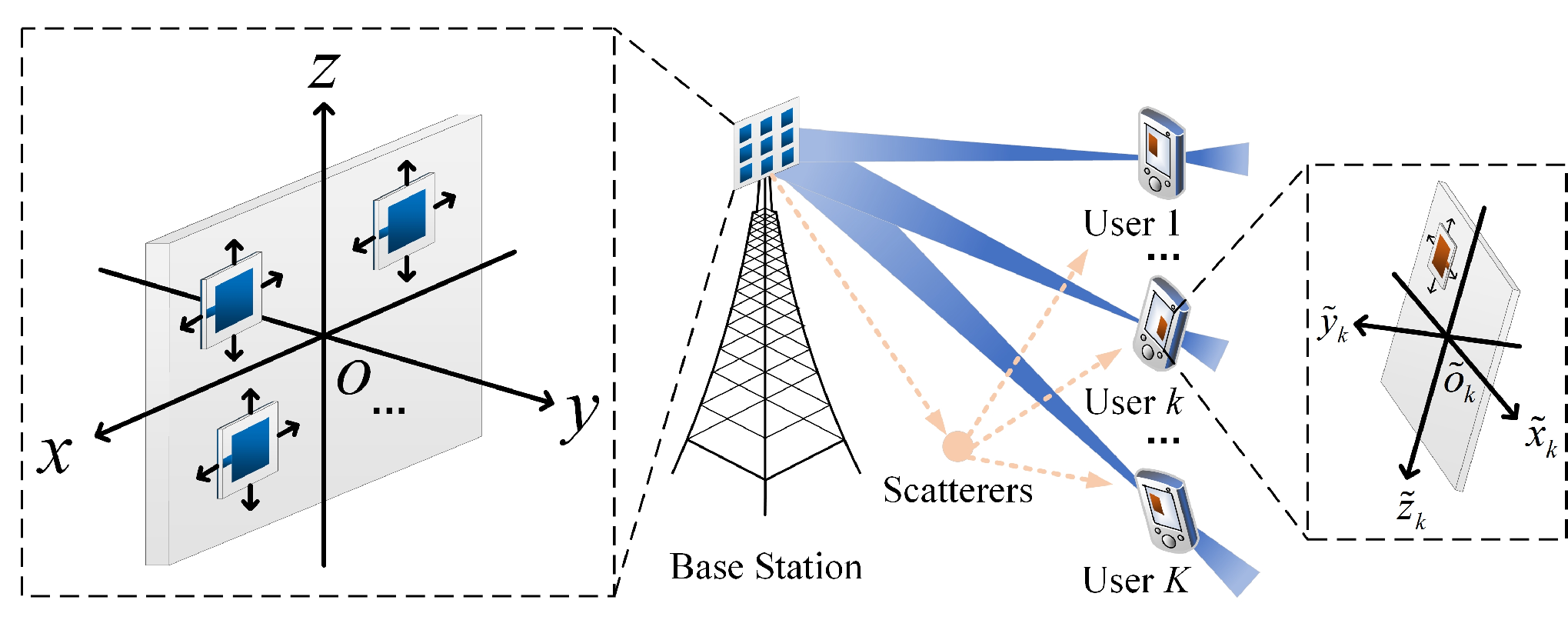}} \\
	\subfloat[]{\label{geometry}\includegraphics[width=1\columnwidth]{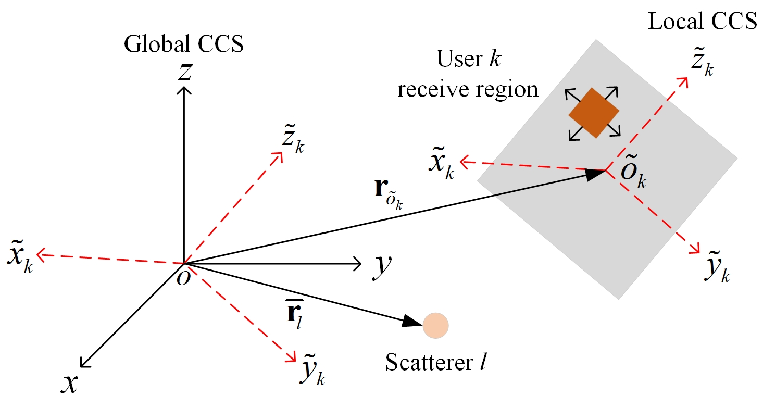}}
	\caption{Illustrations of (a) the proposed system and (b) the geometry of the global and local CCSs.}
	\label{Illustration}
\end{figure}
As illustrated in Fig.\;\ref{Illustration}\subref{system_model}, we examine a DL transmission scenario between the BS and $K$ users. The BS is equipped with $N$ transmit MAs and user $k$ ($k \in \mathcal{K} \triangleq \left\lbrace 1,\cdots,K \right\rbrace $) is equipped with a single receive MA. We define the transmit region as $\mathcal{C}^\mathrm{t}$ and the receive regions of user $k$ as $\mathcal{C}^\mathrm{r}_k$. As shown in Fig.\;\ref{Illustration}\subref{geometry}, we establish a global Cartesian coordinate system (CCS) $o$-$xyz$ at the BS, with the reference point of $\mathcal{C}^\mathrm{t}$ defined as the origin $o$. The global coordinates of $N$ transmit MAs are described by $\mathbf{t} = \left[ {\mathbf{t}_1^T, \cdots ,\mathbf{t}_n^T, \cdots ,\mathbf{t}_N^T} \right]^T \in \mathbb{R}^{3N \times 1}$, where $\mathbf{t}_n = \left[ x^\mathrm{t}_n,y^\mathrm{t}_n,z^\mathrm{t}_n \right]^T \in \mathcal{C}^\mathrm{t}$. We also establish a local CCS ${\tilde o_k}$-${\tilde x_k}{\tilde y_k}{\tilde z_k}$ at user $k$, with the center of $\mathcal{C}^\mathrm{r}_k$ defined as the origin ${\tilde o_k}$. The local coordinates of $K$ receive MAs can be described by $\tilde{\mathbf{r}} = \left[ {\tilde{\mathbf{r}}_1^T, \cdots ,\tilde{\mathbf{r}}_k^T, \cdots ,\tilde{\mathbf{r}}_K^T} \right]^T \in \mathbb{R}^{3K \times 1}$, where $\tilde{\mathbf{r}}_k = \left[ \tilde{x}^\mathrm{r}_k,\tilde{y}^\mathrm{r}_k,\tilde{z}^\mathrm{r}_k \right]^T \in \mathcal{C}^\mathrm{r}_k$. We define the coordinate transform matrix between the global CCS and the local CCS at user $k$ as $\mathbf{R}_k \in \mathbb{R}^{3\times 3}$, where $\mathbf{R}_k$ is a predetermined orthogonal matrix\footnote{$\mathbf{R}_k$ is determined by the orientation of the user's device, which is assumed to be known in this letter. In practice, device orientation can be estimated by various methods such as maximum-likelihood and tensor decomposition \cite{rotation}.} that satisfies $\mathbf{R}_k\mathbf{R}_k^T=\mathbf{I}_3$ \cite{rotation}. Then, by denoting the coordinate of ${\tilde o_k}$ in the global CCS as $\mathbf{r}_{\tilde o_k}$, the MA position of user $k$ in the global CCS can be expressed as
\begin{equation}
	{\mathbf{r}_k} = \mathbf{r}_{\tilde o_k} + {\mathbf{R}_k}{\tilde {\mathbf{r}}_k}.
\end{equation}

In near-field communications, scatterers in the environment can cause multipath propagation, where the users receive signals from both line-of-sight (LoS) and non-LoS(NLoS) paths. Thus, we define the Rician K-factor $\kappa$ to denote the power ratio between the LoS and NLoS components. We consider a quasi-static\footnote{The proposed MA-aided system is designed for slowly varying channels with static or low-mobility terminals \cite{MA2}. For the scenarios with fast-fading channels, the MA positions need to be designed based on the statistical channel state information, which is an important topic for future research.} near-field channel model adopting the point scatterer assumption \cite{near2}. Let $L_k$ denote the total number of scatterers for user $k$ and $\bar{\mathbf{r}}_l$ denote the coordinate of scatterer $l$ ($l \in \mathcal{L}_k \triangleq \left\lbrace 1,\cdots,L_k \right\rbrace $) in the global CCS. The channel vector between the transmit MAs at the BS and the receive MA at user $k$, denoted by $\mathbf{h}_k\in \mathbb{C}^{N \times 1}$, can be expressed as\footnote{Note that the general channel model can also be applied to far-field scenarios straightforwardly.}
\begin{equation}\label{h1}
	{\mathbf{h}_k} = \sqrt {\frac{\kappa }{{\kappa  + 1}}} \breve{\mathbf{h}}_k + \sqrt {\frac{1 }{{\kappa  + 1}}} \bar{\mathbf{h}}_{k},
\end{equation}
where $\breve{\mathbf{h}}_k \triangleq {\rho _k}\mathbf{a}\left( {\mathbf{t},{\mathbf{r}_k}} \right) \in \mathbb{C}^{N \times 1}$ is the LoS component of user $k$ with the channel gain of ${\rho _k}$ and $\bar{\mathbf{h}}_{k} \triangleq  \sum\nolimits_{l \in \mathcal{L}_k}  {\varsigma _l}\bar{\mathbf{h}}_l^\mathrm{I}\left( {\mathbf{t},{{\bar{\mathbf{r}}}_l}} \right)\bar{h}_{l,k}^\mathrm{II}\left( {\bar{\mathbf{r}}_l,{\mathbf{r}_k}} \right)$ is the NLoS component generated by scatterers. Here, ${\varsigma _l}$ represents the complex reflection coefficient of scatterer $l$, $\bar{\mathbf{h}}_l^\mathrm{I}\left( {\mathbf{t},{{\bar{\mathbf{r}}}_l}} \right) = {{{\bar \rho }_l^\mathrm{I}}\mathbf{a}\left( {\mathbf{t},{{\bar{\mathbf{r}}}_l}} \right)} \in \mathbb{C}^{N \times 1}$ is the channel component between the BS and scatterer $l$ with the channel gain of ${\bar \rho }_l^\mathrm{I}$, and $\bar{h}_{l,k}^\mathrm{II}\left( {\bar{\mathbf{r}}_{l},{\mathbf{r}_k}} \right) = {\bar \rho }_{l,k}^\mathrm{II}{{e^{ - \mathrm{j}\frac{{2\pi }}{\lambda }{{\left\| {{{\bar{\mathbf{r}}}_l} - {\mathbf{r}_k}} \right\|}_2}}}}$ is the channel coefficient between scatterer $l$ and user $k$ with the channel gain of ${\bar \rho }_{l,k}^\mathrm{II}$, where $\lambda$ is the carrier wavelength. Additionally, the near-field steering vector between the transmit MAs and position $\mathbf{x}$, i.e., $\mathbf{a}\left( {\mathbf{t},\mathbf{x}} \right) \in \mathbb{C}^{N \times 1}$, where $ \mathbf{x} \in \left\{ {{{\left\{ {{\mathbf{r}_k}} \right\}}_{k \in \mathcal{K}}},{{\left\{ {{{\bar{\mathbf{r}}}_l}} \right\}}_{l \in \mathcal{L}}}} \right\}$, can be calculated as 
\begin{equation}
	\mathbf{a}\left( {\mathbf{t},\mathbf{x}} \right) = \left[ {{e^{ - \mathrm{j}\frac{{2\pi }}{\lambda }{{\left\| {{\mathbf{t}_1} - \mathbf{x}} \right\|}_2}}}, \cdots ,{e^{ - \mathrm{j}\frac{{2\pi }}{\lambda }{{\left\| {{\mathbf{t}_N} - \mathbf{x}} \right\|}_2}}}} \right]^T .
\end{equation}

For a given time slot, the BS transmits $K$ independent information streams to the $K$ users, i.e., $\sum\nolimits_{k \in \mathcal{K}} {{\mathbf{w}_k}{s_k}}$, where ${s_k}$ denotes the transmit signal with normalized power and $\mathbf{w}_k \in \mathbb{C}^{N \times 1}$ is the corresponding beamformer. Therefore, the achievable rate of user $k$ is given by $R_k = \log_2\left(1+ \Gamma _k\right) $ in bits per second per Hertz (bps/Hz), where $\Gamma _k$ is the receive signal-to-interference-plus-noise ratio (SINR) and defined as
\begin{equation}
	{\Gamma _k} = \frac{{{{\left| {\mathbf{h}_k^H{\mathbf{w}_k}} \right|}^2}}}{{\sum\limits_{i \in \mathcal{K}\backslash \left\{ k \right\}} {{{\left| {\mathbf{h}_k^H{\mathbf{w}_i}} \right|}^2}}  + \sigma _k^2}},
\end{equation}
where $\sigma _k^2$ is the average noise power of user $k$.

In this letter, we aim to minimize the transmit power at the BS by jointly optimizing the beamformers and the positions of the transmit and receive MAs. Mathematically, the optimization problem is formulated as
\begin{align} \label{max1}
	& \mathop {\mathrm{minimize} }\limits_{\mathbf{w}_k,\mathbf{t},\tilde{\mathbf{r}}} \quad \sum\limits_{k \in \mathcal{K}} {\left\| {{\mathbf{w}_k}} \right\|_2^2}  \\
	&\mathrm{s.t.} \quad \text{C1}: R_k \ge R^\mathrm{th}_k, \quad \forall k \in \mathcal{K}, \nonumber\\
	&\hspace{2.3em} \text{C2}: \mathbf{t} \in {\mathcal{C}^\mathrm{t}}, \quad \tilde {\mathbf{r}}_k \in {\mathcal{C}^\mathrm{r}_k},\quad \forall k \in \mathcal{K}, \nonumber\\
	&\hspace{2.3em} \text{C3}: {\left\| {{\mathbf{t}_n} - {\mathbf{t}_{\tilde n}}} \right\|_2} \ge d_\mathrm{M} ,\quad 1 \le n \ne {\tilde n} \le N. \nonumber
\end{align}
Here, constraint C1 guarantees the minimum-achievable-rate requirement, $R^\mathrm{th}_k$, for each user. Constraint C2 restrains the restricted moving regions. Constraint C3 ensures the minimum inter-MA distance, $d_\mathrm{M}$, at the BS for practical implementation. Note that the non-convex constraints C1 and C3 and the coupling between the optimization variables render the problem particularly intractable. Thus, we develop suboptimal solutions in the next section.
\section{Proposed Solution} \label{3}
Due to the different types of optimization variables that are intricately coupled, i.e., MA positions and beamformers, conventional alternating optimization (AO) may converge to undesired local optimal solutions because the MA positions (or beamformers) obtained in the previous iteration restrict the optimization space for the beamformers (or MA positions) in the current iteration \cite{Ding2}. To avoid the undesired local optimal solutions, we improve the swarm-intelligence-based standard particle swarm optimization (PSO) algorithm, which is suited for solving multimodal and multivariable optimization problems \cite{Ding1,Ding2}, and propose two-loop DNNPSO algorithm. In the inner-loop, given the MA positions determined by each particle, the beamformers can be optimized separately. In the outer-loop, the position of each particle represents a feasible solution for the MA positions, and the corresponding optimized beamformers are used to evaluate the fitness value of each particle to guide the optimization of the MA positions. The details are presented below.
\subsection{Two-Loop DNPPSO}
First, we define the number of particles as $P$ and randomly initialize the velocity and position of each particle as $\mathbf{v}_p^{\left( 0 \right)} \in {\mathbb{R}^{3\left( N+K\right)  \times 1}}$ and $\mathbf{u}_p^{\left( 0 \right)} = {\left[ { {\mathbf{t}}^{\left(0 \right) T}_p},{{\tilde {\mathbf{r}}}^{\left(0 \right) T}_p} \right]^T} \in {\mathbb{R}^{3\left( N+K\right)  \times 1}}$ ($1 \le p \le P$), respectively\footnote{Note that each element in $\mathbf{u}_p^{\left( 0 \right)}$ is initialized to satisfy constraint C2.}.  Then, we initialize the personal best position of particle $p$, $\mathbf{u}_{\mathrm{pbest},p}$, as $\mathbf{u}_p^{\left( 0 \right)}$, and select the global best position, $\mathbf{u}_{\mathrm{gbest}}$, based on the fitness function under the assumption that the best position has the minimum fitness value. Let $Q$ denote the maximum number of iterations. After completing the initialization, the processing procedures of the two-loop DNPPSO algorithm are summarized in Algorithm\;\ref{DNPPSO}.
\begin{algorithm}[!t]
	\caption{Two-Loop Algorithm for Solving Problem \eqref{max1}}
	\label{DNPPSO}
	\footnotesize
	\renewcommand{\algorithmicrequire}{\textbf{Input:}}
	\renewcommand{\algorithmicensure}{\textbf{Output:}}
	\begin{algorithmic}[1]
		\REQUIRE $N$, $K$, $\lambda$, $\left\lbrace \sigma^2_k\right\rbrace $, $\mathcal{C}^\mathrm{t}$, $\left\lbrace \mathcal{C}^\mathrm{r}_k \right\rbrace $, $\left\lbrace \mathbf{R}_k \right\rbrace $, $\left\lbrace \mathbf{r}_{\tilde o_k}\right\rbrace $,$\left\lbrace L_k\right\rbrace $, $\left\lbrace \bar{\mathbf{r}}_{l}\right\rbrace $, $\left\lbrace R^\mathrm{th}_k \right\rbrace $, $d_\mathrm{M}$, $\left\lbrace {{{\widetilde P}^{\left( q \right)}}}\right\rbrace $, $Q$, $\tau$, $\omega_\mathrm{min}$, $\omega_\mathrm{max}$, $c_1$, $c_2$.
		\ENSURE $\mathbf{u}_\mathrm{gbest}$ and $\mathbf{w}_k$.
		\STATE Initialize the velocity and position of each particles as $\mathbf{v}_p^{\left( 0 \right)}$ and $\mathbf{u}_p^{\left( 0 \right)}$, respectively;
		\STATE Initialize the fitness value of each particle by solving problem \eqref{max2} and update it according to \eqref{fitness};
		\STATE Initialize the personal best position $\mathbf{u}_{\mathrm{pbest},p} = \mathbf{u}_p^{(0)}$ and the global best position $\mathbf{u}_\mathrm{gbest} = \arg \mathop {\min }\limits_{\mathbf{u}_p^{(0)}} \left\{  {\mathcal{F}\left( {\mathbf{u}_1^{(0)}} \right), \cdots, \mathcal{F}\left( {\mathbf{u}_P^{(0)}} \right)} \right\}$;
		\FOR{$q=1:1:Q$}
		\STATE Update the inertia weight $\omega^{\left(q \right)} $ and the number of residual particles ${{{\widetilde P}^{\left( q \right)}}}$;
		\FOR{$p=1:1:{{{\widetilde P}^{\left( q \right)}}}$}
		\STATE Update the velocity and position of the $p$-th particle by \eqref{v} and \eqref{u}, respectively;
		\STATE Calculate the fitness value of each particle by solving problem \eqref{max2} and update it according to \eqref{fitness};
		\IF{$\mathcal{F}\left( {\mathbf{u}_p^{(q)}} \right) < \mathcal{F}\left({\mathbf{u}_{\mathrm{pbest},p}}\right)$}
		\STATE Update $\mathbf{u}_{\mathrm{pbest},p} = \mathbf{u}_p^{(q)}$;
		\ENDIF
		\IF{$\mathcal{F}\left( {\mathbf{u}_p^{(q)}} \right) < \mathcal{F}\left({\mathbf{u}_\mathrm{gbest}}\right)$}
		\STATE Update $\mathbf{u}_\mathrm{gbest} = {\mathbf{u}_p^{(q)}}$;
		\ENDIF
		\ENDFOR
		\STATE Identify the particles within the neighborhood set and remove them;
		\ENDFOR
		\RETURN $\mathbf{u}_\mathrm{gbest}$ and $\mathbf{w}_k$.
	\end{algorithmic}
\end{algorithm}
\subsubsection{Define Fitness Function}
Given the objective to minimize the transmit power, we define the fitness function as
\begin{equation} \label{fitness}
	\mathcal{F}\left( {\mathbf{u}_p^{\left( q \right)}} \right) = {\mathcal{W}}\left( {\mathbf{u}_p^{\left( q \right)}} \right) + \mathcal{P}\left({\mathbf{t}}_p^{\left(q \right)} \right),
\end{equation}
where ${\mathbf{u}_p^{\left( q \right)}}$ is the position of particle $p$ in the $q$-th ($q \in \mathcal{Q} \triangleq \left\lbrace 1,\cdots,Q \right\rbrace $) iteration. In this formulation, $\mathcal{W}\left( \mathbf{u}_p^{\left( q \right)} \right)$ denotes the minimum transmit power derived from solving the following problem for any given MA positions, $\mathbf{u}_p^{\left( q \right)}$:
\begin{equation} \label{max2}
	\mathop {\mathrm{minimize} }\limits_{\mathbf{w}_k} \quad \sum\limits_{k \in \mathcal{K}} {\left\| {{\mathbf{w}_k}} \right\|_2^2} \qquad \mathrm{s.t.} \quad \text{C1}.
\end{equation}
Problem \eqref{max2} is a classical transmit beamformer optimization problem for MU-MIMO systems, which can be efficiently solved via second-order cone programming (SOCP) \cite{SOCP}.

On the other hand, $\mathcal{P}\left( {\mathbf{t}}_p^{\left(q \right)} \right) = \tau \delta\left({\mathbf{t}}_p^{\left(q \right)} \right) $ is the penalty function designed to ensure consistent satisfaction of constraint C3. Here, $\tau$ is a large positive penalty factor and $\delta\left({\mathbf{t}}_p^{\left(q \right)}\right)$ is a counting function that returns the number of transmit MAs violating the minimum inter-MA distance at position ${\mathbf{t}}_p^{\left(q \right)}$. In other words, since we assume that the minimum fitness value corresponds to the best position, the penalty function pushes the particles to satisfy constraint C3.
\subsubsection{Update Positions and Velocities}
The velocity and position of each particle in the $q$-th iteration are updated as
\begin{align}
	&\mathbf{v}_p^{\left( q \right)}  = \omega^{\left( q\right)}  \mathbf{v}_p^{\left( q-1 \right)} + {c_1}{\mathbf{e}_1} \odot \left( {{\mathbf{u}_{\mathrm{pbest},p}} - \mathbf{u}_p^{\left( {q - 1} \right)}} \right) \nonumber\\
	&\hspace{7.5em} + {c_2}{\mathbf{e}_2} \odot \left( {{\mathbf{u}_\mathrm{gbest}} - \mathbf{u}_p^{\left( {q - 1} \right)}} \right) , \label{v} \\
	& \mathbf{u}_p^{\left( q \right)} = \mathcal{B}\left\{ {\mathbf{u}_p^{\left( {q - 1} \right)} + \mathbf{v}_p^{\left( q \right)}} \right\}, \label{u}
\end{align}	
where $\omega^{\left( q\right)}$ is a linear inertia weight function that decreases with the number of iterations within the interval $\left[\omega_\mathrm{min},\omega_\mathrm{max} \right] $, i.e., $\omega = \omega_\mathrm{max}- \left(  \omega_\mathrm{max} - \omega_\mathrm{min}\right) q/Q$. $c_1$ and $c_2$ are the personal and global learning factors, respectively. $\mathbf{e}_1$ and $\mathbf{e}_2$ are random vectors, with each entry uniformly distributed within the range $\left[0,1\right]$. The function $\mathcal{B}\left\lbrace \mathbf{u} \right\rbrace =\max \left\{ {\min \left\{ {\mathbf{u},{\mathcal{C}_\mathrm{up}}} \right\},{\mathcal{C}_\mathrm{low}}} \right\} $ projects each element of the vector $\mathbf{u}$ back into the feasible region if it exceeds the upper bound, $\mathcal{C}_\mathrm{up}$, or the lower bound, $\mathcal{C}_\mathrm{low}$, to guarantee constraint C2.
\subsubsection{Dynamic Neighborhood Pruning}
Once the positions of the particles are determined, the personal and global best positions are updated if the fitness value at the current position is lower than the respective personal and global minimum fitness values. Then, the algorithm initiates the pruning process.

On one hand, calculating the minimum transmit power for each particle's position during each iteration incurs high computational overhead. On the other hand, the particles that are close to the global best position have largely exhausted the potential for discovering better positions. As such, we respectively define the dynamic neighborhood radius and neighborhood set of the global best position as
\begin{align}
	&\left\{ {\left. {d_\mathrm{R}^{\left( q \right)}} \right|\sum\limits_{p = 1}^{{{\tilde P}^{\left( {q} \right)}}} {\xi \left( {{{\left\| {\mathbf{u}_p^{\left( q \right)} - {\mathbf{u}_\mathrm{gbest}}} \right\|}_2} < d_\mathrm{R}^{\left( q \right)}} \right)}  = {{\tilde P}^{\left( {q} \right)}} - {{\tilde P}^{\left( q+1 \right)}}} \right\} , \\
	&\mathcal{N}\left( d_\mathrm{R}^{\left( q \right)} \right) = \left\{ {\left. {\mathbf{u}_p^{\left( q \right)}} \right|{{\left\| {\mathbf{u}_p^{\left( q \right)} - {\mathbf{u}_\mathrm{gbest}}} \right\|}_2} < d_\mathrm{R}^{\left(q \right)},1 \le p \le {\tilde P}^{\left( {q} \right)}} \right\},
\end{align}
where $\xi\left( \cdot \right) $ is an indicator function, equaling one when the condition within the bracket is true; otherwise, it equals zero. ${\tilde P}^{\left( q \right)}$ denotes the predetermined number of residual particles in the $q$-th iteration. In each iteration, the particles within $\mathcal{N}\left( d_\mathrm{R}^{\left( q \right)} \right)$ are pruned to reduce the computational overhead.
\subsection{Convergence and Complexity Analysis}\label{conv_com}
Since only the position with a lower fitness value is selected as the global best position, the global best fitness value is non-increasing during the iterations. Besides, the transmit power is lower-bounded by zero. Therefore, the convergence of the proposed algorithm is guaranteed.

The computational complexity of Algorithm\;\ref{DNPPSO} primarily stems from the iterations of the two-loop DNPPSO algorithm and the process of solving problem \eqref{max2} using SOCP for each particle, for which the complexities are $\mathcal{O} \left( \sum\nolimits_{k \in \mathcal{K}} {\left( {{L_k} + 1} \right)} \sum\nolimits_{q \in \mathcal{Q}} {{{\widetilde P}^{\left( q \right)}}} \right) $ and $\mathcal{O}\left( K^{3.5}N^{3}\right)$ \cite{SOCP}, respectively. Hence, the overall complexity is $\mathcal{O}\left( K^{3.5}N^{3} \sum\nolimits_{k \in \mathcal{K}} {\left( {{L_k} + 1} \right)}\sum\nolimits_{q \in \mathcal{Q}} {{{\widetilde P}^{\left( q \right)}}}\right)$.
\section{Simulation Results} \label{4}
This section presents simulation results to validate the performance of the proposed scheme. In the simulation, the LoS and NLoS near-field channels $\breve{\mathbf{h}}_k$ and $\mathbf{\bar{h}}_{k}$ follow the uniform spherical wave (USW) channel model presented in \cite{near2}, where the complex reflection coefficient ${\varsigma _l}$ is modeled as the complex Gaussian distribution $\mathcal{CN}\left(0,1\right)  $. Without loss of generality, we adopt the linear pruning strategy throughout the iterations, where the number of particles linearly decreases from the initial $P$ to $\beta P$ with $0 < \beta \le 1$. Therefore, the computational complexity is reduced by $\frac{{0.5Q\left( {P - \beta P} \right)}}{{QP}} = 0.5\left( {1 - \beta } \right)$ compared to the standard PSO algorithm.  Here, the linear pruning ratio, $\beta$, is set to 0.02. The transmit and receive regions are set as square areas with sizes $A^\mathrm{t} \times A^\mathrm{t}$ and $A^\mathrm{r}_k \times A^\mathrm{r}_k$, respectively. The carrier frequency is set to 28 GHz ($\lambda=1.07$ cm). The users and scatterers are uniformly distributed around the BS at distances ranging from 50 to 200 m, within the near-field region. Unless otherwise stated, for the proposed communication system, we set the moving region sizes $A^\mathrm{t} = 100\lambda$ and $A^\mathrm{r}_k = \lambda$, the numbers of transmit antennas, users, and scatterers $N=10$, $K=6$, and $L=10$, the average noise power $\sigma _k^2 = -80$ dBm, the Ricean K-factor $\kappa = 3$ dB, and the minimum inter-MA distance $d_\mathrm{M} = \frac{\lambda }{2}$. For the proposed PSO-based algorithm, the parameter settings are based on existing works \cite{Ding1,Ding2}, i.e., the maximum numbers of iterations and particles $Q = 50$ and $P=50$, the personal and global learning factors $c_1 =1.4$ and $ c_2 = 1.4$, the minimum and maximum inertia weights $\omega_\mathrm{min} = 0.4$ and $\omega_\mathrm{max} = 0.9$, and the penalty factor $\tau = 100$.
\begin{figure}
	\centering
	\includegraphics[width=0.7\linewidth]{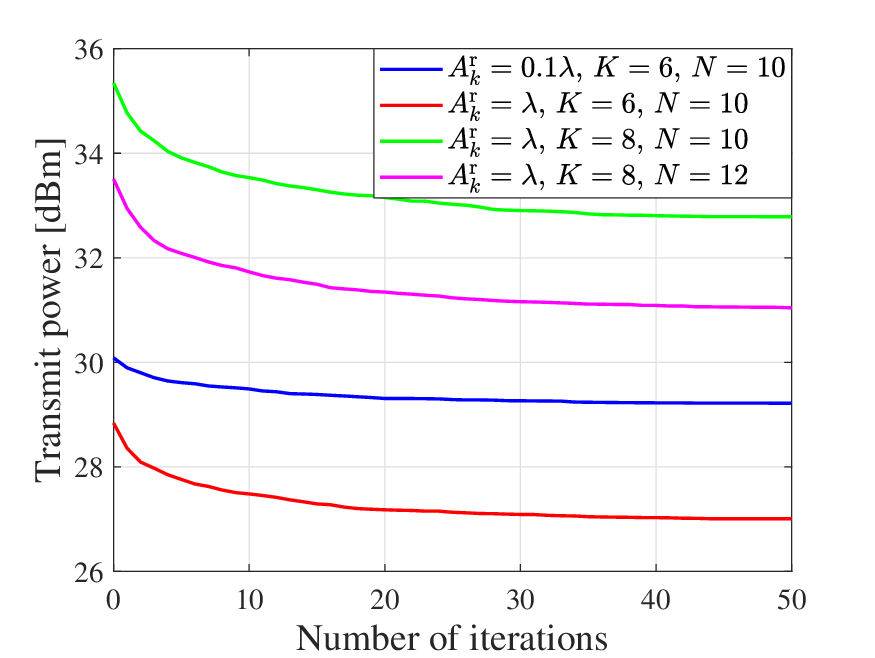}
	\caption{Evaluation of the convergence of the proposed algorithm.}
	\label{conv}
\end{figure}
\begin{table*}[!t]
	\caption{Comparisons of Complexity and Performance}
	\label{tab1}
	\centering
	\begin{tabular}{|c|c|c|c|}
		\hline
		\multicolumn{1}{|c|}{\textbf{Algorithm}} & \multicolumn{1}{c|}{\textbf{Complexity}} & \multicolumn{1}{c|}{\textbf{CPU run time [s]}} &
		\multicolumn{1}{c|}{\textbf{Transmit power [dBm]}} \\ \hline
		AO & $\mathcal{O}\left( I\left( K^{3.5}N^{3}+\left( NG_1+KG_2\right) \sum\nolimits_{k \in \mathcal{K}} {\left( {{L_k} + 1} \right)}\right) \right)$ & 135.06 & 11.28 \\ \hline
		Standard PSO & $\mathcal{O}\left( K^{3.5}N^{3} \sum\nolimits_{k \in \mathcal{K}} {\left( {{L_k} + 1} \right)}PQ\right)$  & 477.19 & 9.71 \\ \hline
		Proposed & $\mathcal{O}\left( K^{3.5}N^{3} \sum\nolimits_{k \in \mathcal{K}} {\left( {{L_k} + 1} \right)}\sum\nolimits_{q \in \mathcal{Q}} {{{\widetilde P}^{\left( q \right)}}}\right)$ & 249.31 & 9.95 \\ \hline
	\end{tabular} 
\end{table*}
\begin{figure*}
	\centering
	\subfloat[]{\label{A_UE}\includegraphics[width=0.5\columnwidth]{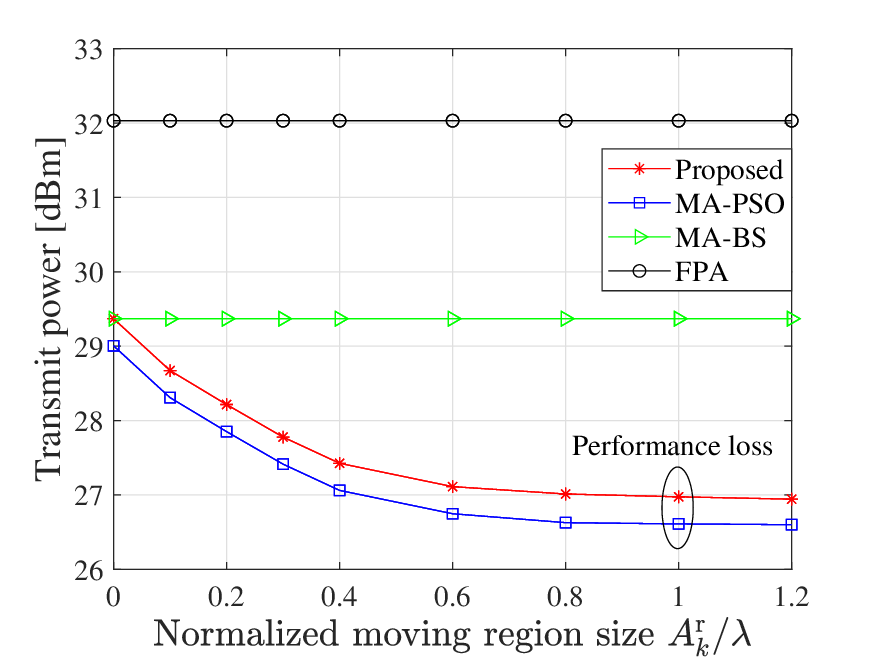}}
	\subfloat[]{\label{K}\includegraphics[width=0.5\columnwidth]{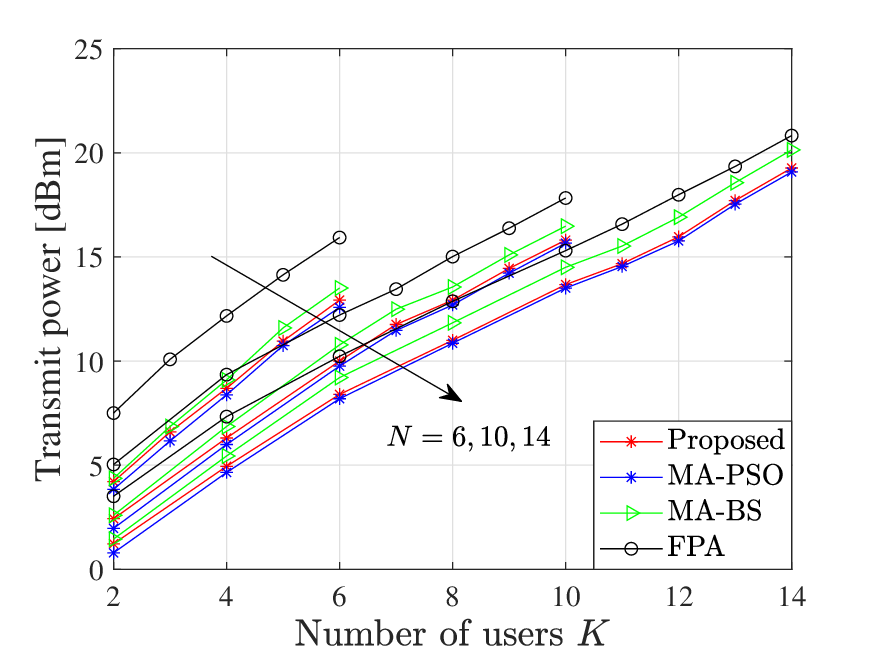}}
	\subfloat[]{\label{R}\includegraphics[width=0.5\columnwidth]{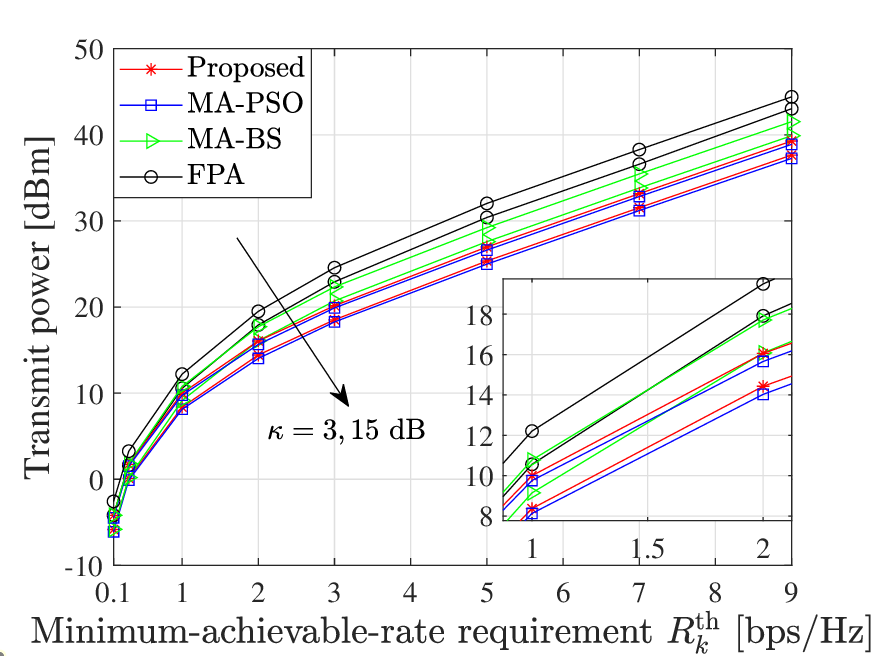}}
	\subfloat[]{\label{distance}\includegraphics[width=0.5\columnwidth]{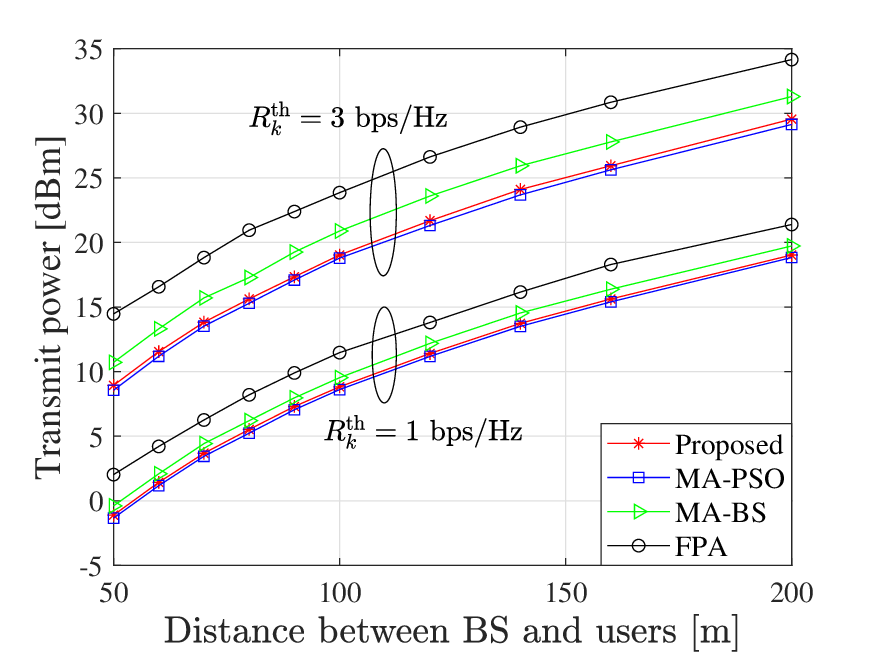}}\\
	\caption{Transmit power versus (a) normalized moving region size, (b) number of users, (c) minimum-achievable-rate requirement, and (d) distance between BS and users.}
	\label{transmit_power}
\end{figure*}

First, Fig.\;\ref{conv} illustrates the convergence behavior of the proposed algorithm with $R^\mathrm{th}_k=5$ bps/Hz. The results show a decrease in transmit power with increasing iterations, which stabilizes within 50 iterations across all configurations. This validates the convergence analysis in Section \ref{conv_com}.

Then, the comparisons of complexity and performance for the proposed and state-of-the-art algorithms are summarized in Table \ref{tab1}\footnote{The AO algorithm follows \cite{Ding1}, where $I$ is the number of AO iterations, and $G_1$ and $G_2$ are the numbers of discretization grids for the transmit and receive regions, respectively. The simulations are performed on MATLAB R2022b using a 64-bit Linux server with 100 virtual cores and 256 GB RAM. The processing speed of each core is 2.6 GHz.} with $R^\mathrm{th}_k=1$ bps/Hz. We can see that the AO algorithm has a low complexity but converges to undesired local optimal solutions, while the standard PSO algorithm finds better solutions at the cost of a higher computational complexity. The proposed algorithm reduces the central processing unit (CPU) run time by 47.75\% with only a 2.47\% performance loss compared to the standard PSO algorithm, striking a balance between computational efficiency and solution quality.

Subsequently, to fully demonstrate the advantages of the proposed scheme, we define the following benchmark schemes: (1) \textbf{MA-PSO}: The MA positions are optimized using the standard PSO algorithm, which is equivalent to setting $\beta =1$. (2) \textbf{MA-BS}: The MAs are only equipped on the BS, and thus we have $\mathbf{u}_p^{\left( q \right)} =  { {\mathbf{t}}^{\left(q \right)}_p}  \in {\mathbb{R}^{3N  \times 1}}$ while ${{\tilde {\mathbf{r}}}^{\left(q \right)}_p} = \mathbf{0}_{3K}$  in the proposed algorithm. (3) \textbf{FPA}: The BS is equipped with a horizontal uniform linear array (ULA) with $N$ antennas spaced by $\frac{\lambda}{2}$, and the users are equipped with a single FPA.

Fig.\;\ref{transmit_power}\subref{A_UE} depicts the transmit power versus the normalized moving region size at the users with $R^\mathrm{th}_k=5$ bps/Hz. The transmit powers of the proposed and MA-PSO schemes decrease with the normalized size of the moving region. This is because expanding the moving region sizes of users' MAs enables them to utilize spatial DoFs more effectively for signal reception. Besides, the negligible performance gap between these two schemes implies that the proposed pruning strategy successfully reduces redundancy in the standard PSO algorithm and the proposed algorithm achieves the precise antenna positioning within a wavelength, which is crucial for antenna movements in near-field scenarios due to the short wavelengths of high-frequency signals. Moreover, the MA-BS and FPA schemes maintain high transmit power levels, as neither scheme implements MAs at both the BS and the users.

Fig.\;\ref{transmit_power}\subref{K} presents the transmit power versus the number of users with $R^\mathrm{th}_k=1$ bps/Hz. As the number of users increases, the transmit powers of different schemes increase due to the need to manage the multiuser interference and meet the minimum-achievable-rate requirement for each user. Compared to the FPA scheme, the MA-BS scheme partially reduces power expenditure, and the proposed scheme further reduces the power consumption by evolving the users' antennas into MAs. As the number of users grows, the gap between the two schemes progressively widens. This indicates that a greater number of MAs can better exploit the DoFs across various terminals and mitigate the severe path loss of high-frequency signals in near-field communications by actively reconfiguring the channels.

Fig.\;\ref{transmit_power}\subref{R} shows the transmit power versus the minimum-achievable-rate requirement for each user. Intuitively, the BS has to transmit signals with higher power to meet the increased minimum-achievable-rate requirements for all users. Additionally, the schemes exhibit lower transmit power at $\kappa=15$ dB compared to $\kappa=3$ dB. This is because the longer propagation distance of NLoS paths compared to LoS paths leads to higher path loss, especially in high-frequency bands. This conclusion is also validated in Fig.\;\ref{transmit_power}\subref{distance}, i.e., longer distance leads to higher path loss. Nevertheless, the proposed scheme still performs well due to the DoFs in antenna movement, allowing flexible beamfocusing to serve users under near-field channel conditions.
\section{conclusion}\label{5}
This letter investigated an MA-aided DL multiuser system under the near-field channel condition, where both the BS and the users are equipped with MAs. We proposed a general channel model to accurately describe the channel characteristics and developed a two-loop DNPPSO algorithm to jointly optimize the beamformers and the MA positions. Simulation results showed that the proposed scheme achieves the precise antenna positioning with a low complexity and reduces the power consumption compared to state-of-the-art schemes.

\newpage

\vfill

\end{document}